\documentclass[%
reprint,
superscriptaddress,
showkeys,
 amsmath,amssymb,
 aps,
]{revtex4-2}
\usepackage{color}
\usepackage{amsfonts}
\usepackage{graphicx}
\usepackage{dcolumn}
\usepackage{bm}
\usepackage{lineno}

\begin{document}

\preprint{APS/123-QED}

\title{Optimized architectures for universal quantum state transformations \\ using photonic path and polarization }


\author{Dong-Xu Chen}
 \email{chendx@sru.edu.cn}
 \affiliation{Quantum Information Research Center, Shangrao Normal University, Shangrao, Jiangxi 334001, China}
\author{Junliang Jia}%
\affiliation{Shaanxi Key Laboratory of Quantum Information and Quantum Optoelectronic Devices, School of Physics of Xi'an Jiaotong University, Xi'an 710049, China}
\author{Pei Zhang}%
 \email{zhang.pei@xjtu.edu.cn}
\affiliation{Shaanxi Key Laboratory of Quantum Information and Quantum Optoelectronic Devices, School of Physics of Xi'an Jiaotong University, Xi'an 710049, China}
\author{Chui-Ping Yang}%
 \email{yangcp@hznu.edu.cn}
 \affiliation{Quantum Information Research Center, Shangrao Normal University, Shangrao, Jiangxi 334001, China}
 \affiliation{School of Physics, Hangzhou Normal University, Hangzhou, Zhejiang 311121, China}

\date{\today}
\begin{abstract}
An arbitrary lossless transformation in high-dimensional quantum space can be decomposed into elementary operations which are easy to implement, and an effective decomposition algorithm is important for constructing high-dimensional systems. Here, we present two optimized architectures to effectively realize an arbitrary unitary transformation by using the photonic path and polarization based on the existing decomposition algorithm. In the first architecture, the number of required interferometers is reduced by half compared with previous works. In the second architecture, by using the high-dimensional X gate, all the elementary operations are transferred to the operations which act locally on the photonic polarization in the same path. Such an architecture could be of significance in polarization-based applications. Both architectures maintain the symmetric layout. Our work facilitates the optical implementation of high-dimensional transformations and could have potential applications in high-dimensional quantum computation and quantum communication.
\end{abstract}

\maketitle

\section{Introduction}
A lossless transformation in high-dimensional (HD) quantum space maps $N$ inputs into $N$ outputs, where $N$ is the dimension of the input quantum states. HD transformations on quantum states are important in various quantum information task, e.g., quantum computation \cite{knill2001scheme}, quantum simulation \cite{spring2013boson,univer,harris2017quantum} and quantum metrology \cite{PhysRevLett.114.170802}. The realization of HD transformations generally requires decomposing them into elementary operations which are easy to implement experimentally.

The issue about decomposition of HD transformations for photonic path was resolved by Reck \textit{et al.} in their seminal work \cite{PhysRevLett.73.58}, where HD transformation matrices are decomposed into interferometers operating between two adjacent modes. The design forms a triangular layout. Then it was improved by Clements \textit{et al.} \cite{Clements:16} with a rectangular design which is symmetric with respect to different paths. The symmetric structure yields the robustness against the noise introduced by the interferometers, and a shorter optical path. While in principle, the decomposition of HD transformations into elementary operations applies to any degree of freedom, the linear optics is one of the most promising platforms to realize HD transformations. Besides, the photonic path is a natural carrier to encode HD quantum information and it is experimentally mature to realize the operations between two adjacent paths. On the other hand, the integrated photonic circuit is becoming a promising platform to realize linear optical experiments compared with the bulk optical elements, because of its reconfigurability, high stability, and versatility \cite{bogaerts2020programmable,9205209,chen2021,lu2021,al2022}. Therefore, the design in \cite{Clements:16} has become a primary route for implementing HD transformations in integrated photonic circuits.

Despite the explicit decomposition of HD transformations in \cite{Clements:16}, some new designs were proposed to further simplify the practical realization. For example, an algorithm that decomposed HD transformations into a sequence of HD transformations of one dimension less, entangled by a beam splitter, was proposed in \cite{PhysRevA.97.022328}. Hybrid spatiotemporal architectures were proposed to combine the benefits of the spatial and temporal degrees of freedom of a photon \cite{PhysRevA.99.062301}. In \cite{PhysRevLett.124.010501}, the authors proposed a robust architecture based on multichannel blocks. Also, in \cite{PhysRevA.92.043813}, the authors demonstrated that by using the photonic path and internal degrees of freedom, the number of required interferometers can be reduced, at the cost of more elements acting on the internal degrees of freedom. 

In this work, based on the existing decomposition algorithm, we present two optimized architectures to realize HD transformations in the hybrid HD space by combining the path and the polarization of a photon. In the first architecture, the elementary operations are realized by the rotating operations on the polarization and the path of a photon. Since the operations on the polarization can be realized by a combination of wave plates without interferometer structures, the number of required interferometers is reduced by half compared with \cite{Clements:16}.

In the second architecture, by using the HD X gate and X$^{\dagger}$ gate, we transfer all the elementary operations to the rotating operations which act only on the polarization of the photon in the same path. As a consequence, beam splitters and phase shifters are not needed, and only optical elements acting on the polarization are required. Meanwhile, since the interferometers are replaced by wave plates, the photon only interferes at the X gate and X$^{\dagger}$ gate. Therefore, the optical depth is reduced by half compared with \cite{Clements:16}. Both architectures maintain the symmetric structure in \cite{Clements:16}, thus preserving the robustness against the noise.

It is advantageous to encode information with the photonic internal degrees of freedom, since the quantum states of the internal degrees of freedom could maintain mutual stability. It was shown in \cite{d2012complete} that using the photonic internal degrees of freedom could allow alignment-free quantum communication, since the phases of the states in the hybrid space, introduced by the rotation of the reference frame, were cancelled. Also, each degree of freedom can be used to encode information, hence an HD quantum space can be constructed by coupling multiple degrees of freedom of a photon. Related works include using multiple degrees of freedom of a photon to generate HD entangled states \cite{PhysRevLett.120.260502}, to realize quantum teleportation between multiple degrees of freedom \cite{wang2015quantum}, \textcolor{black}{to implement HD quantum teleportation \cite{PhysRevLett.123.070505,PhysRevLett.125.230501}}, and to construct HD quantum gates \cite{Wang_2021,PhysRevA.103.022606}.

We remark that the polarization of a photon is a powerful and widely used degree of freedom to encode information in both classical and quantum applications. Though existing researches on integrated photonic circuits are mainly focused on photonic path, attention has been paid to the on-chip manipulation of photonic polarization \cite{PhysRevLett.105.200503,pitsios2017geometrically,10.1117/12.2251407,zeuner2018integrated}. Meanwhile, coherent conversion of photonic quantum entanglement between multiple degrees of freedom has been realized \cite{feng2016chip}. Therefore, to realize HD transformations in path-polarization hybrid space could be an important step towards on-chip HD quantum computation.

\section{Elimination-based decomposition}
First, we briefly review the rectangular design proposed by Clements \textit{et al.} in \cite{Clements:16}. The algorithm factors an arbitrary HD unitary transformation matrix $U_N$ into a product of a diagonal matrix and a sequence of splitting operation matrices
\begin{equation}
U_N=D_N\left( \prod_{(m,n)\in S} T_{m,n}\right),
\label{Eq.UN}
\end{equation}
where $S$ defines a specific order that the matrices $T_{m,n}$ operate on the input state; $D_N$, which corresponds to an HD phase gate, is a diagonal matrix with complex elements with a modulus equal to one on the diagonal; $T_{m,n}$ is the splitting operation between two adjacent modes $m$ and $n$ ($n=m+1$), and it has the following form
\begin{eqnarray}
T_{m,n}&=&\left[\begin{array}{cccccccc}
1 & 0& \cdots &\cdots &\cdots &\cdots &\cdots & 0\\
0 & 1& &&&& & \vdots\\
\vdots & & \ddots&&&& & \vdots\\
\vdots & & &e^{i\phi}\cos\theta & -\sin\theta & & & \vdots\\
\vdots & & &e^{i\phi}\sin\theta & \cos\theta & & & \vdots\\
\vdots & & & & &\ddots & & \vdots\\
\vdots & & & & & &1 & 0\\
0 &  \cdots &\cdots &\cdots &\cdots &\cdots &0& 1\\
\end{array}\right],
\end{eqnarray}
i.e., it is the identity matrix with the entries at the intersection of the $m$th and $n$th rows and columns replaced by the following matrix
\begin{eqnarray}
R(\theta,\phi)&=&\left[\begin{array}{cc}
e^{i\phi}\cos\theta & -\sin\theta \\
e^{i\phi}\sin\theta & \cos\theta\\
\end{array}\right].
\end{eqnarray}
Note that the $\theta$ and $\phi$ of each $T_{m,n}$ in Eq.~(\ref{Eq.UN}) are not necessarily equal. For simplicity, we write $T_{m,n}$, without explicitly indicating the parameters $\theta$ and $\phi$, to represent a general splitting operation between the $m$th mode and the $n$th mode. Here we only focus on the matrices $T_{m,n}$ and omit the phase gate, since it can be realized by inserting phase shifters in each mode in the end.

\begin{figure}[tbp!]
\centering\includegraphics[width=0.45\textwidth]{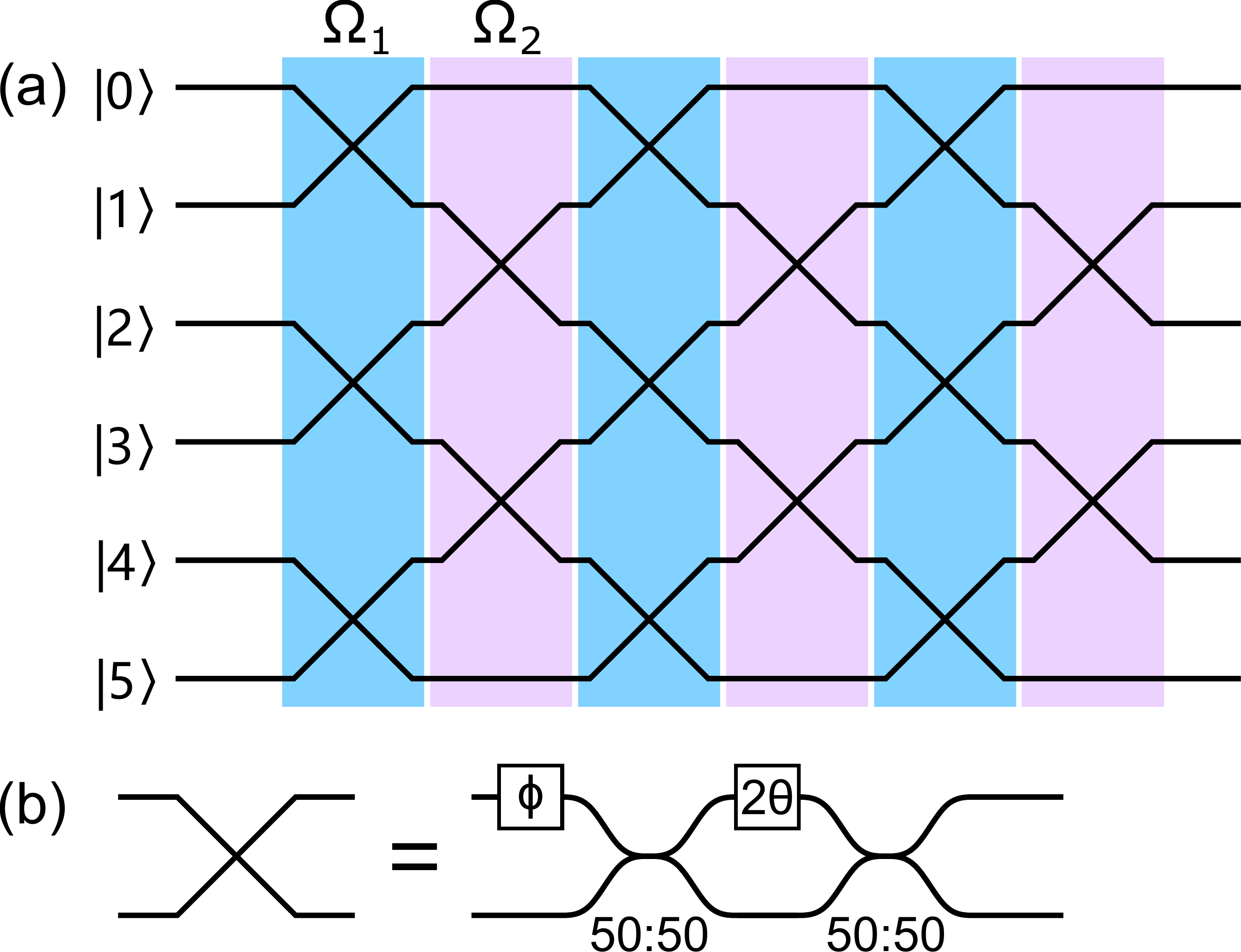}
\caption{(a) A $6\times 6$ unitary transformation can be decomposed into 15 interferometers. Each black line represents one input mode. The crossing between two lines represents one interferometer. The operations in each $\Omega_1$ layer (blue region) and each $\Omega_2$ layer (purple region) act on the quantum states alternately. (b) An interferometer is composed of two balanced beam splitters and two tunable phase shifters.}
\label{fig1}
\end{figure}

For photonics paths, each $T_{m,n}$ can be realized by a Mach-Zehnder interferometer (MZI), which is composed of two balanced beam splitters and two phase shifters. The number of required interferometers scales as $N(N-1)/2$ \cite{Clements:16}. Figure \ref{fig1}a shows an example of decomposing a 6 $\times$ 6 unitary transformation $U_{6}$ into 15 interferometers. The input modes are encoded with the logical states $|0\rangle, |1\rangle, |2\rangle, |3\rangle, |4\rangle$ and $|5\rangle$, respectively, which are represented by black solid lines. Each crossing between two modes represents an interferometer, shown in Fig.~\ref{fig1}b.

From the layout of Fig.~\ref{fig1}a, one can see that the rectangular design in \cite{Clements:16} can be regarded as containing two types of layers. One layer, denoted by $\Omega_1$ in the blue region in Fig.~\ref{fig1}a, contains the splitting operations between the $j$th mode and the $(j+1)$th mode, where $j$ is an even number; while the other layer, denoted by $\Omega_2$ in the purple region in Fig.~\ref{fig1}a, contains the splitting operations between the $(j-1)$th mode and the $j$th mode. The operations in these two layers act on the quantum state alternately. Each $\Omega_1$ layer contains three different operations, $T_{0,1}$, $T_{2,3}$ and $T_{4,5}$; while each $\Omega_2$ layer contains two different operations, $T_{1,2}$ and $T_{3,4}$. The operations in the same layer commute with each other, which means the operating order is irrelevant in the same layer.

\section{Results}
While the photonic path is a natural candidate to encode HD quantum information, one can also use other degrees of freedom of a photon to encode HD quantum states, such as the time \cite{steinlechner2017distribution,PhysRevLett.118.110501,islam2017provably} and orbital angular momentum \cite{10010,100100}, or combination of multiple degrees of freedom. Thus, a 6-dimensional space can also be formed by two polarization states and three path states. Here, we use two polarization states (denoted by $|h\rangle$ and $|v\rangle$) and three path states (denoted by $|a\rangle$, $|b\rangle$ and $|c\rangle$) to form a 6-dimensional space. We present two optimized architectures to effectively realize an arbitrary unitary transformation in such a hybrid space based on the elimination-based decomposition algorithm in \cite{Clements:16}. In the first architecture, the number of required MZIs is reduced by half since some elementary operations could be replaced by the rotating operations acting on the polarization. It is a hybrid path-polarization operation architecture, which means we need operations acting on both the path and the polarization of a photon. The second architecture is a full polarization operation architecture, where each of the elementary operations is transferred to an operation acting on the polarization in the same path, by using an HD X gate and its inverse operation. Hence, only polarization-related optical elements are required. 

\begin{figure}[htbp]
\centering\includegraphics[width=0.5\textwidth]{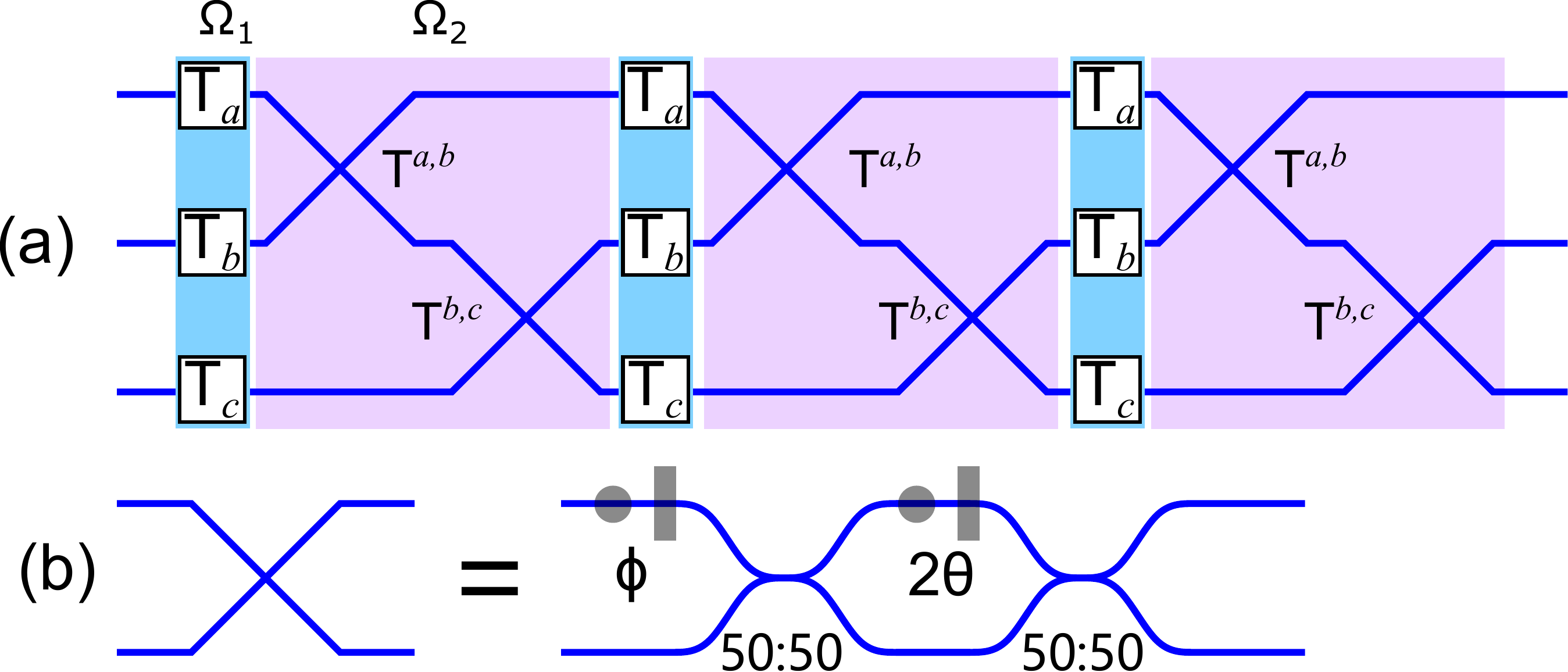}
\caption{(a) Hybrid path-polarization operation architecture. The operations in $\Omega_1$ are realized by polarization rotating operations $T_a$, $T_b$, and $T_c$ in each path. The operations in $\Omega_2$ are realized by two polarization-dependent beam splitters $T^{a,b}$ and $T^{b,c}$. (b) Construction of a polarization-dependent beam splitter. Before the first balanced beam splitter, a phase shifter (represented by a gray circle) and a combination of wave plates (represented by a gray rectangle) introduce phase $\phi$. Here, the phase shifter introduces an overall phase, while the wave plates introduces a relative phase between horizontal and vertical polarizations. The phase $2\theta$ is also set in the same way. Such a polarization-dependent beam splitter can realize different operations on horizontal and vertical polarization states.}
\label{fig2}
\end{figure}

\subsection{Hybrid path-polarization operation architecture}
In the first architecture, the logical states in the 6-dimensional space are encoded in the following order	
\begin{eqnarray}
\nonumber  |0\rangle&\equiv&|a,h\rangle,\quad |1\rangle\equiv|a,v\rangle,\quad |2\rangle\equiv|b,v\rangle,\\
|3\rangle&\equiv& |b,h\rangle,\quad|4\rangle\equiv |c,h\rangle,\quad|5\rangle\equiv |c,v\rangle,
\label{encode1_1}
\end{eqnarray}
\textcolor{black}{i.e., the polarization state follows the order $|h\rangle$, $|v\rangle$, $|v\rangle$, $|h\rangle$, $|h\rangle$, and $|v\rangle$.} Our scheme is shown in Fig.~\ref{fig2}a, where a blue line represents one path containing two perpendicular polarization states, to differentiate from the black line in Fig.~\ref{fig1} which represents one mode. Each crossing between two lines represents a polarization-dependent beam splitter, shown in Fig.~\ref{fig2}b.

Based on the encoding in Eq.~(\ref{encode1_1}), it turns out that each operation in $\Omega_1$ layers is a polarization rotating operation in each path. For example, the operation $T_{0,1}$ mixes the states $|0\rangle$ and $|1\rangle$, which correspond to states $|a,h\rangle$ and $|a,v\rangle$, respectively. Therefore, the operation $T_{0,1}$ can be implemented by a combination of wave plates placed in path $a$. In general, the operations $T_{0,1}$, $T_{2,3}$, and $T_{4,5}$ in $\Omega_1$ can be implemented by $T_a$, $T_b$, and $T_c$, respectively. Here, $T_p$ is the operation acting on the polarization in path $p$ ($p=a,b,c$). Generally, the operation $T_p$ can be realized by combination of half-wave plates and a quarter-wave plate \cite{PhysRevA.85.022323}.

On the other hand, the operations in $\Omega_2$ mix the states of the same polarization in two adjacent paths. To be specific, $T_{1,2}$ mixes the states $|a,v\rangle$ and $|b,v\rangle$, while $T_{3,4}$ mixes the states $|b,h\rangle$ and $|c,h\rangle$. Note that $T_{1,2}$ only acts on the states with vertical polarization in paths $a$ and $b$, while for states with horizontal polarization, it could be regarded as an identity operator. Similarly, $T_{3,4}$ only acts on the states with horizontal polarization in paths $b$ and $c$.

\begin{figure*}[htbp!]
\centering\includegraphics[width=1\textwidth]{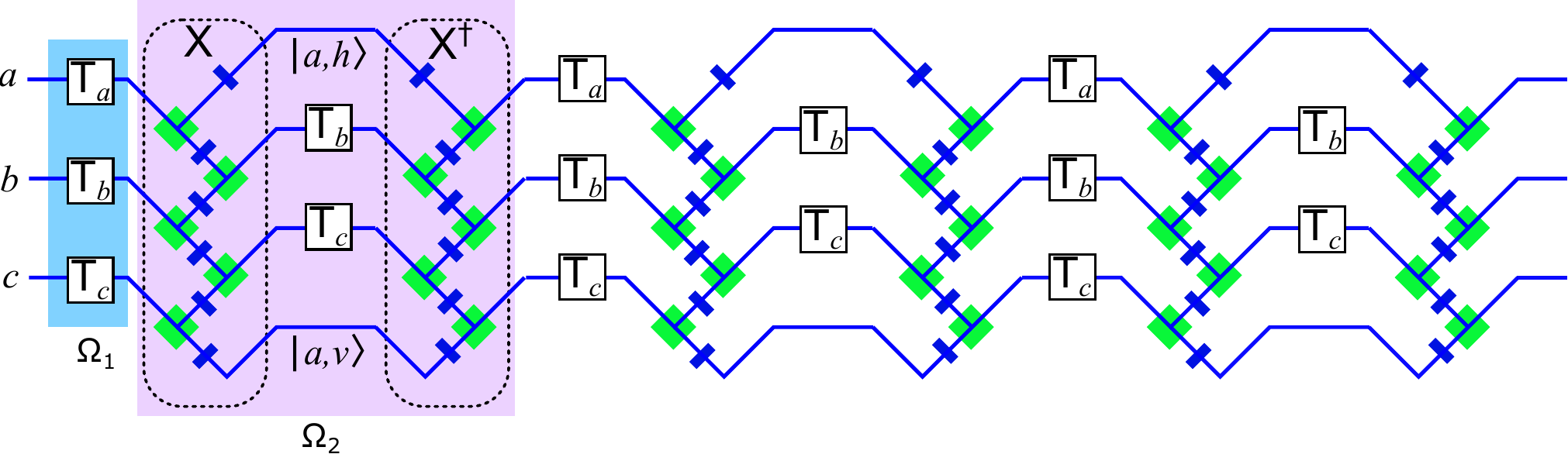}
\caption{Full polarization operation architecture. Each blue line represents one path which contains two polarizations. The operations in $\Omega_1$ can be realized by operations $T_p$ in each path, with $T_p$ being the polarization rotating operation. The operations in $\Omega_2$ can be realized by an HD X gate, followed by $T_b$ and $T_c$, and an HD X$^{\dagger}$ gate. The dashed boxes in $\Omega_2$ are the implementations of HD X gate and X$^{\dagger}$ gate, respectively. Here, the states $|a,h\rangle$ and $|a,v\rangle$ are spatially separated after the X gate. The green squares represent polarization beam splitters which transmit horizontally polarized photons and reflect vertically polarized photons. The blue rectangles represent half-wave plates oriented at $45^{\circ}$.}
\label{fig3}
\end{figure*}

Therefore, one could use polarization-dependent beam splitters (shown in Fig.~\ref{fig2}b) to realize the operations in $\Omega_2$. The polarization-dependent beam splitter is an MZI with the phases $\phi$ and $2\theta$ being polarization-dependent. As shown in Fig.~\ref{fig2}b, the phases $\phi$ and $2\theta$, which are introduced by phase shifters in Fig.~\ref{fig1}, are now introduced by phase shifters and wave plates. Here, the phase shifters introduce overall phases, while the wave plates introduce relative phases between horizontal and vertical polarizations. Hence, such a construction can impose different operations to horizontal and vertical polarization states. For $T_{1,2}$, one could use $T^{a,b}$, which represents a polarization-dependent beam splitter operating on paths $a$ and $b$, to impose the desired operation on the photon with vertical polarization; while it works as an identity operator for the photon with horizontal polarization. Similarly, $T_{3,4}$ can be replaced by $T^{b,c}$.

Then, the operations in $\Omega_1$ and $\Omega_2$ layers are performed on the quantum states alternatively three times, producing the output states. In this architecture, the interferometers in $\Omega_1$ layers are replaced by the operations acting on the polarization, and only the realization of $\Omega_2$ requires MZI structures. Therefore, the number of required interferometers is reduced by half. Both the operations acting on the polarization and the operations acting on the path are needed.

\subsection{Full polarization operation architecture}

In the second architecture, we show that each of the elementary operations can be transferred to an operation acting locally on the polarization states in the same path. For simplicity, the logical states are encoded in the following order
\begin{eqnarray}
\nonumber  |0\rangle&\equiv&|a,v\rangle,\quad |1\rangle\equiv|a,h\rangle,\quad |2\rangle\equiv|b,v\rangle,\\
|3\rangle&\equiv& |b,h\rangle,\quad|4\rangle\equiv |c,v\rangle,\quad|5\rangle\equiv |c,h\rangle,
\label{encode2}
\end{eqnarray}
which is different from the encoding in Eq.~(\ref{encode1_1}). The layout of the second architecture is shown in Fig.~\ref{fig3}. Based on the encoding of the logical states in Eq.~(\ref{encode2}), one can straightforwardly conclude that the operations in $\Omega_1$ layers can similarly be implemented by the operations $T_p$, which are the polarization rotating operations in each path. 

On the other hand, the operations in $\Omega_2$ mix different polarizations in different paths, thus a combination of wave plates is not sufficient. However, they can be regarded as acting on the two states in $\Omega_1$ with each state being transformed into the next state. Therefore, if each state is transformed into the next state, the operations in $\Omega_2$ will effectively act on the polarization states in the same path. For example, the operation $T_{1,2}$ in $\Omega_2$ mixes the states $|a,h\rangle$ and $|b,v\rangle$. If one transforms $|a,h\rangle$ into $|b,v\rangle$, and $|b,v\rangle$ into $|b,h\rangle$, the operation $T_{1,2}$ will be transformed into the operation $T_{2,3}$. Then according to the above, it can be implemented by the operation $T_b$, which is a combination of wave plates in path $b$.

Such a transformation operation on the states is equivalent to the operation described by an HD X gate, which takes the cyclic operation on the HD states. The HD X gate has the following form \cite{PhysRevLett.116.090405,Schlederer_2016,Chen_2017,PhysRevLett.119.180510,isdrailua2019cyclic,PhysRevA.99.023825}
\begin{eqnarray}
X=\sum_{j=0}^{j=N-1}|j\oplus 1\rangle\langle j|,
\end{eqnarray}
where $j\oplus 1\equiv \textrm{mod}(j+1, N)$. Consequently, the operations in $\Omega_2$ can be replaced by
\begin{eqnarray}
T_{1,2}\Rightarrow X^{\dagger}T_bX,\quad T_{3,4}\Rightarrow X^{\dagger}T_cX,
\label{T}
 \end{eqnarray}
which can also be verified through the corresponding matrix multiplication. The implementations of X and X$^{\dagger}$ are shown in the dashed boxes in Fig.~\ref{fig3}. The green squares represent polarization beam splitters which transmit horizontally polarized photons and reflect vertically polarized photons. The blue rectangles represent half-wave plates oriented at $45^{\circ}$. Note that the states $|a,h\rangle$ and $|a,v\rangle$ are spatially separated after the X gate, though they are logically different polarization states in the same path. Such a layout is reasonable since the two states are not transformed in $\Omega_2$. 

One can see from Fig.~\ref{fig3} that each elementary operation is transferred to the operation $T_p$ acting on the polarization in the same path. Therefore, such an architecture only requires polarization operations.

\section{Scaling properties}
Both architectures proposed above can be generalized to the case of arbitrary dimensions. Consider a $2n$-dimensional space which is composed of two polarization states and $n$ path states $|p_1\rangle$, $|p_2\rangle,\cdots,|p_n\rangle$. In this section, we illustrate the generalization of the two architectures and \textcolor{black}{discuss the corresponding scaling properties from three aspects, the optical depth, the number of required interferometers, and the photon loss. The transmission coefficients of a balanced beam splitter, a polarization beam splitter, and a wave plate are denoted as $\eta_{\rm b}, \eta_{\rm p}$, and $\eta_{\rm w}$, respectively. The transmission coefficient of a phase shifter in an MZI shown in Fig. \ref{fig1} is denoted as $\tilde{\eta}_{\rm ph}$. In addition, the transmission coefficient of the elements, which are used to introduce the combined phases in a polarization-dependent beam splitter shown in Fig. \ref{fig2}, is defined as ${\eta}_{\rm ph}$. Generally, we have $\tilde{\eta}_{\rm ph}>{\eta}_{\rm ph}$. Therefore, to realize a $2n\times 2n$ unitary operation, for a path-encoded scheme in \cite{Clements:16}, the overall transmission coefficient is $\eta=(\eta_{\rm b}\tilde{\eta}_{\rm ph})^{4n}$. Here, one MZI consists of two balanced beam splitters and two phase shifters, and the optical depth is $2n$.}

\begin{figure}[tbp!]
\centering\includegraphics[width=0.42\textwidth]{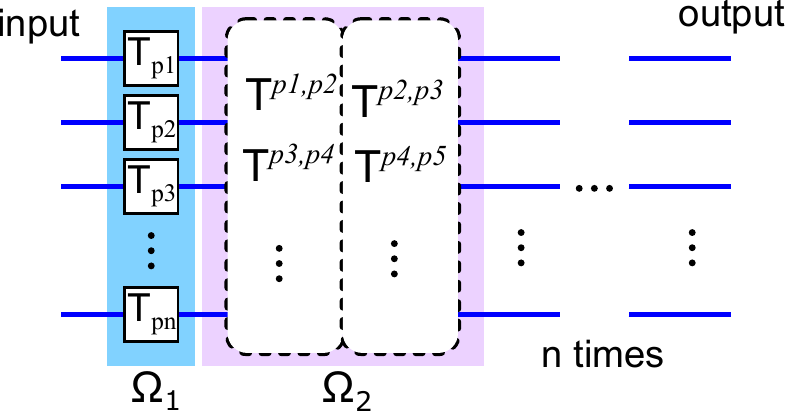}
\caption{Generalization of hybrid path-polarization operation architecture for an arbitrary $2n\times 2n$ unitary transformation. Each $\Omega_1$ layer can be realized by $n$ operations acting on the polarization in each path. Each $\Omega_2$ layer is realized in two steps. The operations in the two layers act on the quantum state for $n$ times.}
\label{fig4}
\end{figure}

\subsection{Hybrid path-polarization operation architecture}
The encoding of the logical states in a $2n$-dimensional space follows the order in Eq.~(\ref{encode1_1}), i.e., two adjacent states either transmit along the same path or possess the same polarization. The generalization of the first architecture is shown in Fig.~\ref{fig4}. 

Here, the operations in the $\Omega_1$ layers are all replaced by polarization rotating operations $T_{p_n}$ in each path. Therefore, the realization of each $\Omega_1$ layer requires $n$ combinations of wave plates. On the other hand, in $\Omega_2$ layers, the realization of each $\Omega_2$ layer is divided into two parts, since each path is mixed with the previous and the next paths. In the first part, polarization-dependent beam splitters are placed to realize operations $T^{p_{j-1},p_{j}}$, where $j$ is an even number; while in the second part, the operations $T^{p_{j},p_{j+1}}$ are realized.

It is obvious that in this case, we need $n^2$ combinations of wave plates for $\Omega_1$ layers. For $\Omega_2$ layers, we need $n(n-1)$ polarization-dependent beam splitters. Meanwhile, since the operations in $\Omega_2$ are realized in two steps, the optical depth of our scheme is $2n$. Therefore, such an architecture has an optical depth which is equal to the one in \cite{Clements:16}, whereas the number of required interferometers is reduced by half.

\textcolor{black}{The transmission rate of each $\Omega_1$ layer is $\eta^3_{\rm w}$, where we consider $T_p$ to be realized by three wave plates. The transmission rate of each $\Omega_2$ layer is $(\eta_{\rm b}\eta_{\rm ph})^4$. Therefore, the overall transmission rate is $\eta_1=[\eta^3_{\rm w}(\eta_{\rm b}\eta_{\rm ph})^4]^n$.}

\subsection{Full polarization operation architecture}
\begin{figure}[tbp!]
\centering\includegraphics[width=0.5\textwidth]{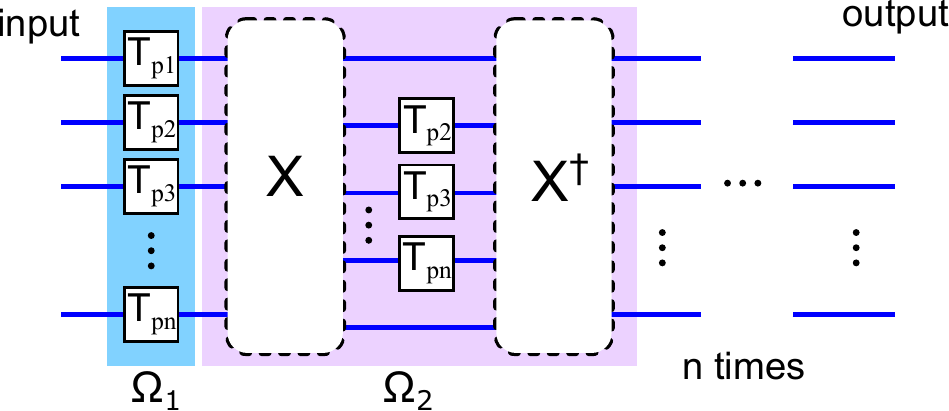}
\caption{Generalization of full polarization operation architecture for an arbitrary $2n\times 2n$ unitary transformation. Each $\Omega_1$ layer can be realized by $n$ operations acting on the polarization in each path, while each $\Omega_2$ layer can be realized by an X gate, $(n-1)$ operations on the polarization in the $(n-1)$ paths, and an X$^{\dagger}$ gate. The operations in the two layers act on the input state for $n$ times. Note that after the X gate, the states $|p_1,h\rangle$ and $|p_1,v\rangle$ are spatially separated, though they logically represent different polarization states in the same path.}
\label{fig5}
\end{figure}

For the generalization of the second architecture, the logical states are encoded following the same order in Eq.~(\ref{encode2}), i.e., the polarization of two adjacent modes is different. The network to implement a $2n$-dimensional transformation using the second architecture is shown in Fig.~\ref{fig5}. Each $\Omega_1$ layer can be implemented by operations $T_{p_n}$ which act locally on the polarization in each path. Then each $\Omega_2$ layer can be implemented by $(n-1)$ polarization operations in $(n-1)$ paths, sandwiched between an X gate and an X$^{\dagger}$ gate. The operations in the two layers are performed on the input state for $n$ times, producing the output state.

Since such a construction transfers each elementary operation to the polarization rotating operation in each path, it does not require phase shifters and beam splitters. Only wave plates and polarization beam splitters are required. As shown in Fig.~\ref{fig5}, each $\Omega_1$ layer contains $n$ operations $T_p$, which are the polarization rotating operations in each path. On the other hand, each $\Omega_2$ layer contains $(n-1)$ operations $T_p$, an X gate, and an X$^{\dagger}$ gate. Each X gate requires $2n$ half-wave plates with fixed angles and $(2n-1)$ polarization beam splitters. The same is for the X$^{\dagger}$ gate. Therefore, the realization of one $\Omega_2$ layer requires $(n-1)$ operations $T_p$, $4n$ half-wave plates with fixed angles, and $2(2n-1)$ polarization beam splitters. Table \ref{table1} summarizes the optical elements required for implementing a $2n$-dimensional transformation using such an architecture.

\begin{table}[htbp!]
\caption{\label{table1}Optical elements required to realize an arbitrary $2n\times 2n$ transformation in the path-polarization hybrid space with a full polarization operation architecture. PBS: polarization beam splitter, HWP(fixed): half-wave plate with fixed angle at $45^{\circ}$, combined wave plates: a combination of wave plates to realize the polarization rotation.}
\begin{ruledtabular}
\begin{tabular}{ccc}
\textrm{PBS}&\textrm{HWP (fixed)} &\textrm{combined wave plates} \\
\colrule
$2n(2n-1)$ & $4n^2$ & $n(2n-1)$ \\
\end{tabular}
\end{ruledtabular}
\end{table}
It is worthy to note that using the design in \cite{Clements:16} requires $n(2n-1)$ MZIs, whereas our scheme requires $n(2n-1)$ combinations of wave plates. This means that the elementary operations acting on the path degree of freedom in \cite{Clements:16} are transferred to the operations acting on the polarization in our scheme, at the cost of addition X gates and X$^{\dagger}$ gates, which are implemented by polarization beam splitters and half-wave plates with fixed orientation. On the other hand, our scheme reduces half of the optical depth, which is defined as the maximum number of interferometers that the photon traverses \cite{Clements:16}. Since the interferometers are replaced by wave plates in each path, these operations do not contribute to the optical depth of the layout. The photon only interferes in the X gate and X$^{\dagger}$ gate. Each $\Omega_2$ layer can be regarded as comprising one interferometer. Therefore, the optical depth for realizing a $2n$-dimensional transformation in our scheme is $n$, which is half of the optical depth in \cite{Clements:16}.

\textcolor{black}{The transmission rate of an X gate is $\eta^2_{\rm p}\eta_{\rm w}$. Therefore, the transmission rate of each $\Omega_2$ layer is $\eta^2_{\rm p}\eta^5_{\rm w}$. Then, the overall transmission rate of the architecture in Fig. \ref{fig5} is $\eta_2=(\eta_{\rm p}\eta^4_{\rm w})^{2n}$.}

\section{Discussion and conclusion}
It is interesting to compare our architectures with the one in \cite{PhysRevA.92.043813}. In \cite{PhysRevA.92.043813}, the authors used path and internal degrees of freedom of a photon to realize an $n_sn_p\times n_sn_p$ unitary transformation, where $n_s$ and $n_p$ are the dimensions of the path and internal degrees of freedom, respectively. They devised an algorithm which reduces the number of beam splitters by a factor $n^2_p/2$ via the cosine-sine (CS) decomposition, at the cost of more optical elements acting the internal modes. For $n_s=n$ and $n_p=2$, the design has an optical depth between $n$ and $2n$. As can be seen from Fig.~4 in \cite{PhysRevA.92.043813}, for the case of $n_s=4$, the decomposition yields 6 CS matrices. The realization of each CS matrix requires an interferometer. The optical depth in this case is 5 since the two CS matrices in the middle are commutable. However, the design does not yield a symmetric layout, which is less noise-resistant compared with symmetric layouts.

We remark that both our architectures maintain the symmetric layout since our designs are based on the decomposition algorithm in \cite{Clements:16}. In our first architecture, half of the interferometers are replaced by combinations of wave plates. Operations on path and polarization are both needed. The optical depth is the same as that in \cite{Clements:16}. In the second architecture, all the elementary operations are transferred to the operations acting on the polarization. Interference of the photon only exists in X gates and X$^{\dagger}$ gates. The number of required interferometers in our scheme is also, to some extend, reduced by half, if one considers the structure formed by an X gate and an X$^{\dagger}$ gate as interferometers. Therefore, this architecture yields a shorter optical depth which equals $n$.


\textcolor{black}{Combining multiple degrees of freedom of a photon is an efficient way to construct an HD quantum space, since it requires less resources and it could also simplify quantum state manipulation in some situations. Based on our schemes, one could also use more internal degrees of freedom of a photon, e.g., frequency and orbital angular momentum, to further simply the realization of HD quantum state transformation. Previous works have been proposed to utilize spatial and temporal degrees of freedom for universal quantum circuits \cite{PhysRevA.99.062301}. The time-multiplexed technique uses discretized arrival times of a photon to encode a qubit, i.e., time-bin encoding. Such technique requires the minimal resources and can in principle be used to construct a high-dimensional space for achieving universal HD quantum state transformations. Theoretical and experimental works on time-bin encoded quantum computation have been presented in recent years \cite{PhysRevLett.113.120501,PhysRevLett.118.190501}. However, the interval of the pulse should be large enough for the detector to effectively distinguish each pulse. Generally, it should be larger than the coherent time of the photon, the dead time of the detector, and the switch time of the optical elements. Then, a larger pulse interval would require a longer delay line, which introduces a photon loss. The photon loss becomes detrimental as the dimension grows. Nevertheless, the advance of modern technology would promote the application of the time-bin encoding in HD quantum information processing.}

\begin{figure}[tbp!]
\centering\includegraphics[width=0.45\textwidth]{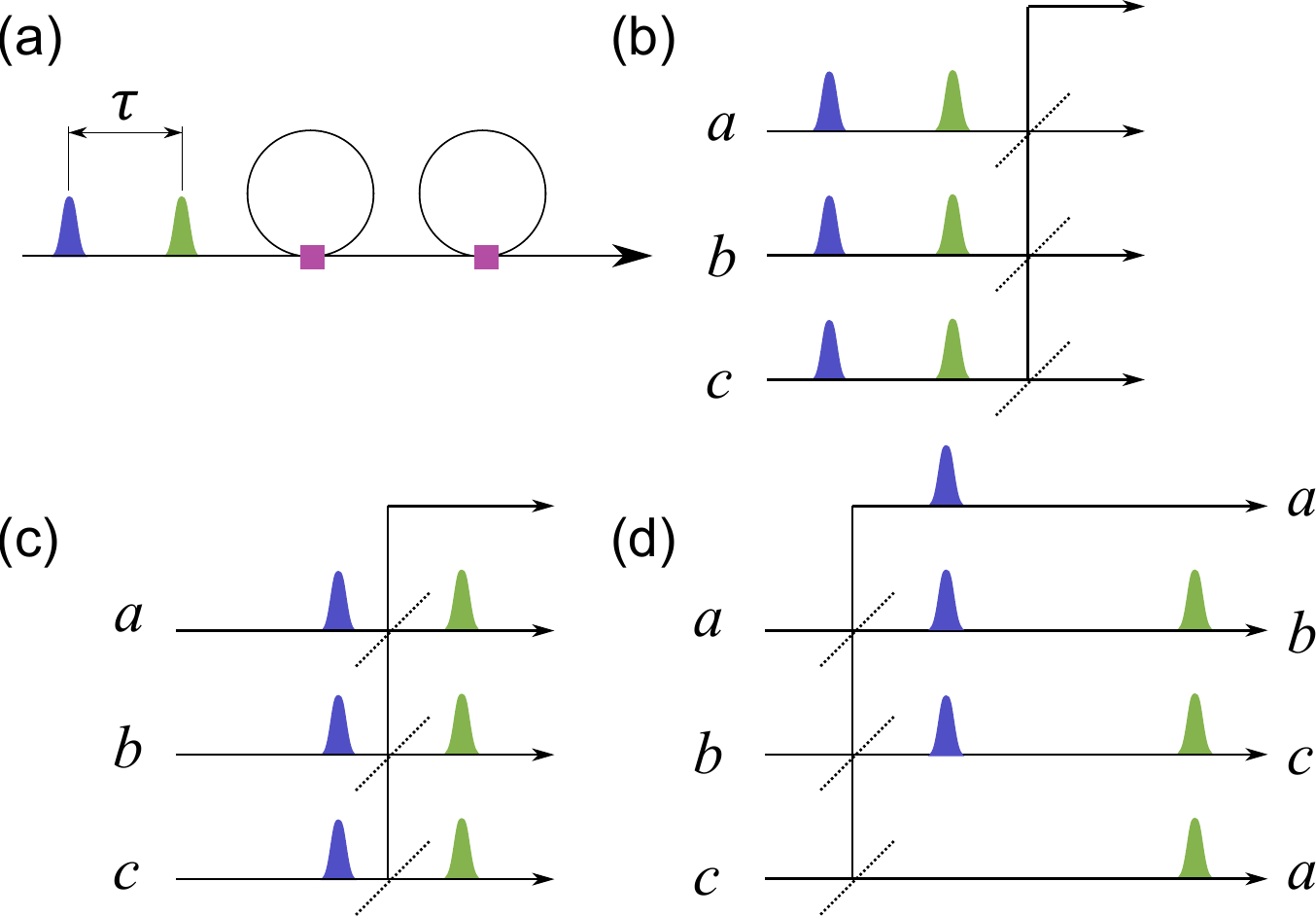}
\caption{(a): Two consecutive loops can realize arbitrary operations on two time bins. The squares represent dynamically controlled beam splitters. (b)-(d): Realization of HD X gate on a 6-dimensional space. The dotted lines represent dynamically controlled mirrors.}
\label{fig6}
\end{figure}

\textcolor{black}{Finally, we remark that polarization-encoded quantum information processing on an integrated photonic chip is currently difficult to achieve, since building polarization-related elements on a chip is hard with the state of the art. Recent works have demonstrated the fabrication of a polarization beam splitter and wave plates on a chip \cite{shen2015integrated,Lu:15,lpr16,lpr19,heilmann2014arbitrary}. Though the realization of the two architectures presented in our work is challenging, they could still be applied in bulk optics with low dimensions to simply the experimental design. Moreover, the ideas presented in our work, i.e., coupling two degrees of freedom of a photon and using X gate, applies to any degrees of freedom of a photon, provided the relevant operations could be realized. For instance, one could use the path and the time bin of a photon to construct the two architectures. Figure \ref{fig6} shows an alternative of using path and time bin encoding to realize universal operations in a 6-dimensional space, which is composed of three paths and two time bins. In this case, the polarization operations $T_p$ correspond to operations on the time bins in the same path. This can be realized by two consecutive loops \cite{PhysRevLett.113.120501}, shown in Fig.~\ref{fig6}a.}

\textcolor{black}{In Fig.~\ref{fig6}a, two pulses with interval $\tau$ represent a photon with different arrival times. The length of the loop is equal to the distance between the pulses. The beam splitter is dynamically controlled to achieve the desired operation. The realization of the HD X gate in such a composite space is shown in Figs. \ref{fig6}(b-c). Each path is equipped with a mirror (Fig.~\ref{fig6}b). The mirror is turned off until the first pulse passes through (Fig.~\ref{fig6}c). Then the mirror is turned on and reflects the latter pulse to the upper path (Fig.~\ref{fig6}d). A swap gate on the time bins in each path is required (not shown in the figure), which can be incorporated into the subsequent operation. Note that the pulses of the output photon in path $a$ is spatially separated. Such a structure is similar to that in Fig.~\ref{fig3} where photon with different polarizations in path $a$ is spatially separated after X gate. Also, the interval between the pulses is increased, but this could be restored after the following X$^{\dagger}$ gate. Combining these two operations, one can realize universal operations with a \textit{full temporal operation} architecture.}

In conclusion, we have presented two optimized architectures to effectively realize an arbitrary unitary HD transformation in a path-polarization hybrid space based on the existing decomposition algorithm. Both architectures reduce the number of required interferometers while maintaining the symmetric layout. Our work facilitates the implementation of HD transformations in a hybrid space and could have potential applications in HD quantum information processing and quantum communication.



\begin{acknowledgments}
This work was supported by the Key-Area Research and Development Program of GuangDong province (2018B030326001), and the National Natural Science Foundation of China (NSFC) (11774076, U21A20436, 12174301, 12204312).
\end{acknowledgments}



\end{document}